\documentclass[journal]{IEEEtranTIE}
\usepackage{setspace}
\usepackage{graphicx}
\usepackage{cite}
\usepackage{picinpar}
\usepackage{amsmath}
\usepackage{url}
\usepackage{flushend}
\usepackage[utf8]{inputenc}
\usepackage{colortbl}
\usepackage{soul}
\usepackage{multirow}
\usepackage{pifont}
\usepackage{color}
\usepackage{alltt}
\usepackage[hidelinks]{hyperref}
\usepackage{enumerate}
\usepackage{siunitx}
\usepackage{breakurl}
\usepackage{epstopdf}
\usepackage{pbox}
\usepackage{tikz}
\usetikzlibrary{shapes.geometric, arrows}

\tikzstyle{startstop} = [rectangle, rounded corners, text centered, draw=black]
\tikzstyle{process} = [rectangle, text centered, draw=black]
\tikzstyle{arrow} = [thick,->,>=stealth]

\begin{document}
\title{
    High-Power Wide-Bandwidth High-Quality Modular Pulse Synthesizer with Adaptive Voltage Asymmetry in Medical Power Electronics
}

\author{
	\vskip 1em
	
	Jinshui Zhang
	and Stefan M.\, Goetz

}

\maketitle
	
\begin{abstract}
    Noninvasive brain stimulation can write signals into neurons but requires power electronics with exceptionally high power in the mega-volt-ampere range and kilohertz usable bandwidth. Whereas oscillator circuits offered only one or very few pulse shapes, modular cascaded power electronics solved a long-standing problem for the first time and enabled arbitrary software-based synthesis of the temporal shape of stimuli. However, synthesizing arbitrary stimuli with a high output quality requires a large number of modules.
    We propose an alternative solution that achieves high-resolution pulse shaping with fewer modules by implementing high-power wide-bandwidth voltage asymmetry. Rather than  equal voltage steps, our system strategically assigns different voltages to each module to achieve a near-exponential improvement in resolution. The module voltage sequence does also not use just a simple binary pattern other work might suggest but adapts it to the output. Additionally, we introduce a switched-capacitor charging mechanism that allows the modules to charge to different voltages through a single dc power supply. 
    We validated our design in a head-to-head comparison with the state of the art on experimental prototypes. 
    Our three-module prototype reduces total voltage distortion by 13.4\% compared to prior art with three modules, and by 4.5\% compared to prior art with six -- twice as many -- modules.
    This paper is the first asymmetric multilevel circuit as a high-precision high-power synthesizer, as well as the first to adaptively optimize asymmetric voltage sequence in modular power electronics.
\end{abstract}


{}

\definecolor{limegreen}{rgb}{0.2, 0.8, 0.2}
\definecolor{forestgreen}{rgb}{0.13, 0.55, 0.13}
\definecolor{greenhtml}{rgb}{0.0, 0.5, 0.0}

\section{Introduction}
\begin{figure}
    \centering
    \includegraphics{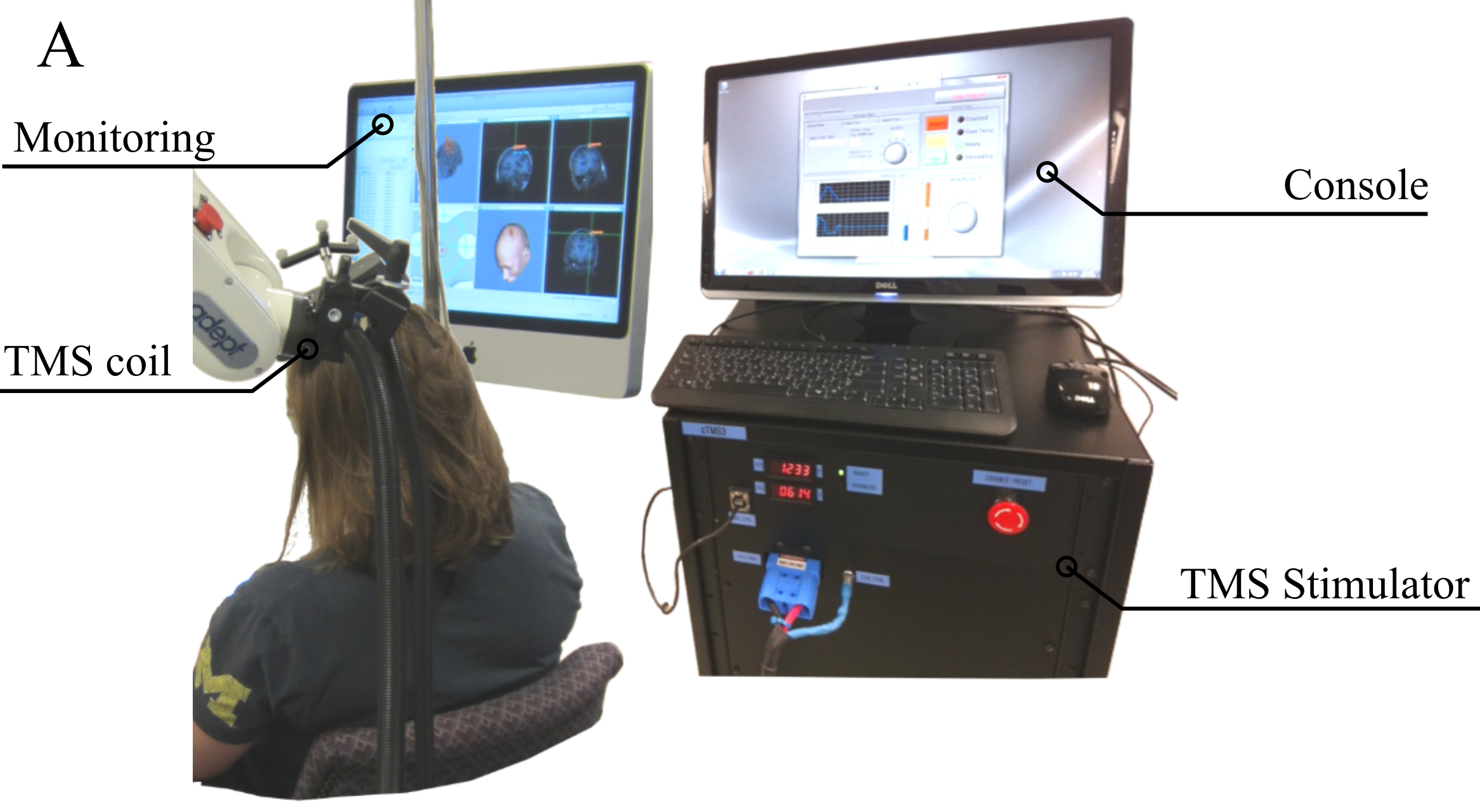}
    \caption{Standard TMS setup consisting of a stimulator circuit, a coil, a console and monitoring equipment.}
    \label{fig:standard_tms_setup}
\end{figure}

Transcranial magnetic stimulation (TMS) is a noninvasive technique that uses very brief powerful magnetic field pulses to induce currents around neurons in the brain, which in turn let electrically sensitive proteins in neurons respond and generate voltage signals in these neurons, which the circuits process very similarly to physiological signals (Fig.\ \ref{fig:standard_tms_setup}) \cite{KLOMJAI2015208}. The power electronics needed for this procedure is extreme compared to more mainstream inverters, e.g., in drives. The current is in the kiloampere range and requires often kilovolts to ramp it up fast enough  \cite{DengTechnology}. The spectral bandwidth of a pulse is in the kilohertz range. At the same time, the output quality particularly of the voltage has to be high. The voltage quality is important as  in the linear range the voltage is proportional to the induced electric field, which is the component that activates neurons. Conventional inverter technology fails as it cannot provide the exceptionally high power, high bandwidth, and high quality at the same time.

However, the brain contains a number of different neuron types and shapes, which differ also in their nonlinear activation dynamics \cite{10.1093/cercor/bhy339}. Different temporal shapes of stimuli can allow selective stimulation \cite{goetz2016enhancement, KAMMER2001250}. This observation stimulated an intensive search in power electronics to find a circuit technology that can synthesize practically any pulse shape \cite{Moritz2002, Talebinejad2016, Peterchev_2011, goetz2013analysis}.

\begin{figure}
    \centering
    \includegraphics{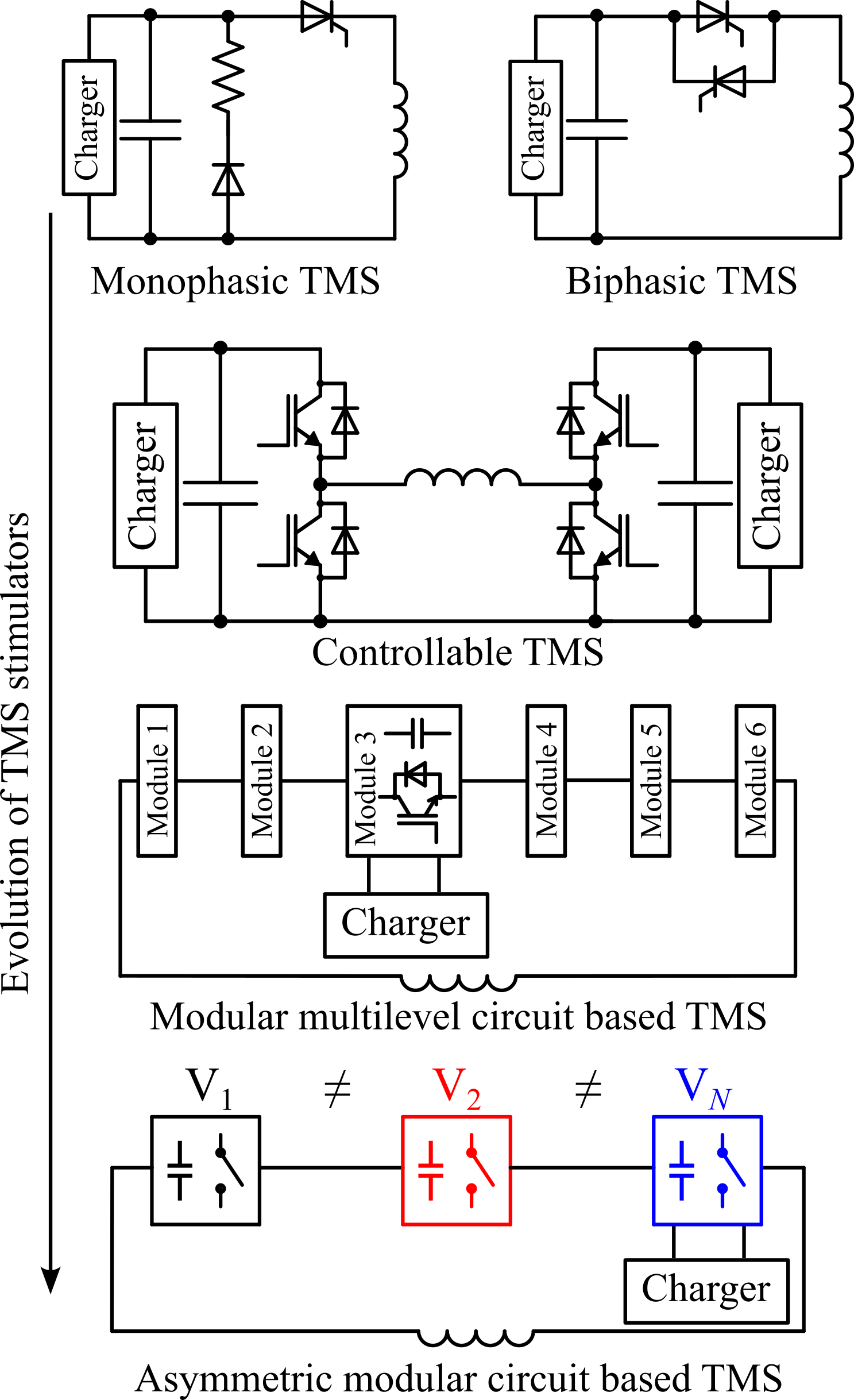}
    \caption{Evolution of TMS pulse generators. Starting from oscillating circuits, the stimulator design embraces more flexibility by introducing switching mode power electronics. The state-of-the-art adopts modular power electronics for a high power scale and high resolution. The latest practices adopt the same voltage for all modules, which requires a large number of modules for a good output quality. The proposed asymmetric multilevel circuit solution obtains a high resolution with differentiating module voltages.}
    \label{fig:evolution_of_tms_pulse_generator}
\end{figure}

Traditional high-power inverters designed for electricity grids are typically optimized for low frequencies, often constrained to the grid frequency. While these systems can handle substantial power levels, their bandwidth remains limited \cite{7884555,9672538,9940564}. Electric drives only increase the bandwidth by approximately one power of ten, which is still far below TMS requirements \cite{6231806, 9135026, 6342749, 6397383}. Resonant circuits -- prevalent in the early stages of TMS circuit evolution (see Figure \ref{fig:evolution_of_tms_pulse_generator}) -- have demonstrated potential for achieving both high power and high output frequency \cite{Wang_2022, sebesta2021sub, 9336630, peterchev2015advances}. However, these circuits are inherently limited by their narrow bandwidth and therefore allow usually only one pulse class.

Modular electronic circuits have resolved this dilemma for the first time. 
The high scalability and flexibility make cascaded multicell circuits the dominant choice for power delivery and conversion applications, including high-voltage AC/DC converters \cite{5597952, 5544594, 5637505, 6038879, 6085335} and medium-voltage motor drives \cite{5984128, 6397383, 5279054, 6342749, 6231806, 8601394}. Moreover, various cascaded circuits can distribute voltage, current, and switching across multiple modules and semiconductors \cite{6824228,9415177}. This approach enables the simultaneous achievement of high bandwidth, high quality, and high power \cite{mps2012circuit}.

In latest TMS power circuits, modular circuits now enable the synthesis of virtually any practical pulse shape. Implementations include modular pulse synthesizers with 300 kHz usable bandwidth at a 10 MVA power level \cite{Li_2022} or 30 kHz bandwidth and 100 MVA power level \cite{Zeng_2022}. However, these machines require a relatively large number of modules to achieve a sufficiently high output quality and low harmonics.

We propose and develop an alternative approach that reduces the number of modules and  still generates a smooth electric field profile with high resolution. Instead of equal voltage steps, our system assigns different operating voltages to each module. Whereas in previous TMS technology the field granularity improved linearly with the number of modules, our approach achieves a near-exponential growth  in resolution. Cascaded converters have previously used asymmetric voltage distributions \cite{6908214, 9070653, 7987947, 5680638, 10758434}. 
Most asymmetric multilevel converters follow a binary or tertiary voltage distribution. Some exceptions shrinked the difference range for a better practicality \cite{10758434}.
However, this paper goes beyond a simple revisit of these ideas. Instead, it develops a customizable voltage asymmetry through optimized designs in topology, hardware, and algorithms.

\section{Module Design}
This section illustrates the design of individual modules, including their topology and design of key components. 

\subsection{Topology}
\begin{figure}
    \centering
    \includegraphics[width=0.45\textwidth]{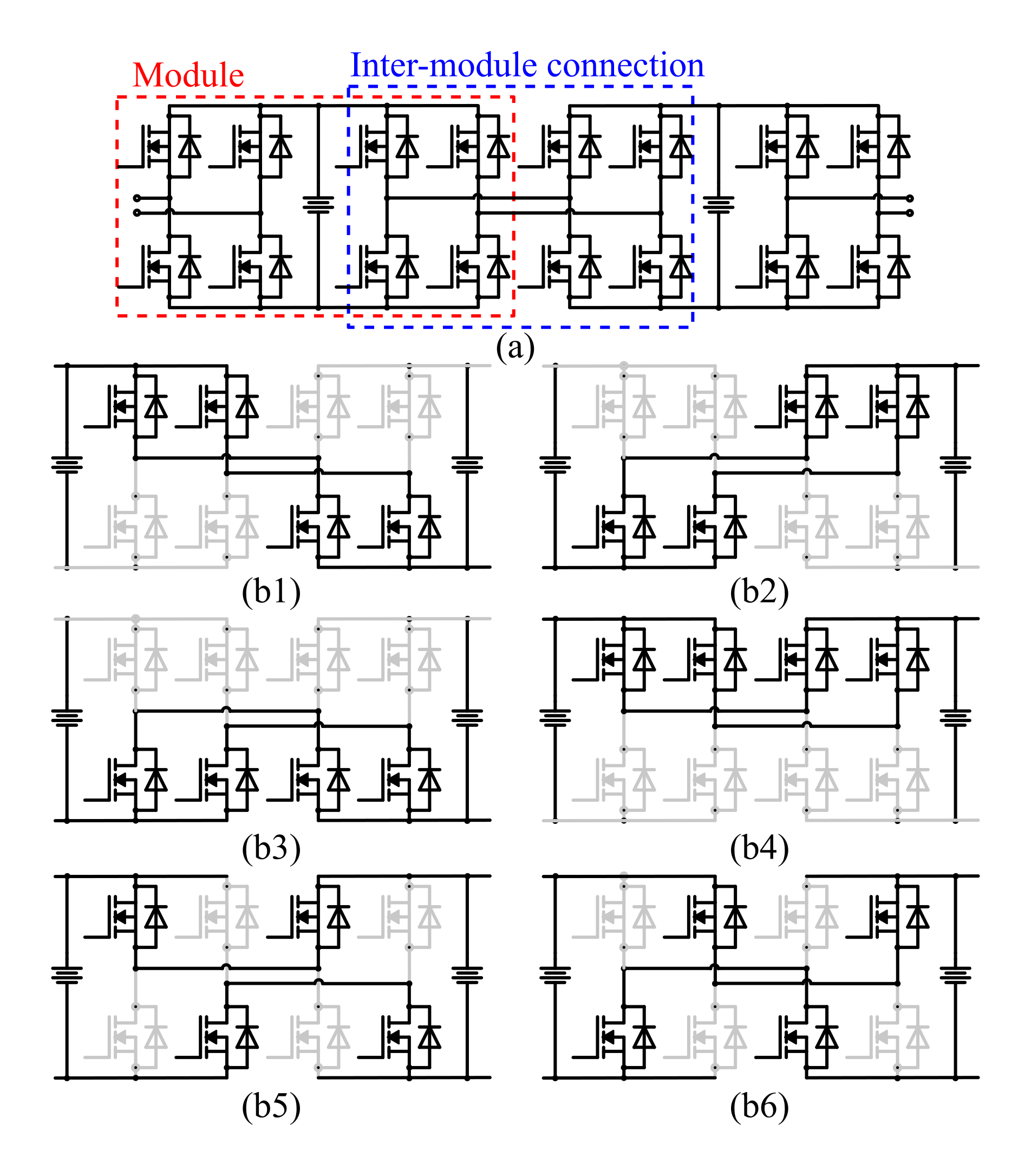}
    \caption{Structure and working principle of cascaded double h-bridge modules. (a) Module topology and inter-module connection. (b) Different switching modes of inter-module connection, including \textit{series--} (b1), \textit{series+} (b2), \textit{bypass--} (b3), \textit{bypass+} (b4), \textit{parallel--} (b5) and \textit{parallel+} (b6).}
    \label{fig:ch2b_principle}
\end{figure}
We adopted cascaded double H-bridge (CH2B) as the topology for all modules \cite{6763109}. Figure \ref{fig:ch2b_principle} illustrates the structure (a) and working principles (b1--b6) of this topology. Each module comprises four half bridges in parallel and four terminals to connect with other modules. The module status are controlled with the basic unit of inter-module connection, which supports six states of \textit{Bypass+, Bypass--, Series+, Series--, Parallel+} and \textit{Parallel--}. These states respectively generate the output of \{0, 0, +V, --V, 0, 0\}. While the parallel configuration is usually desired for reduced impedance \cite{10509104}, we leverage this feature to  charge modules to different voltages using a single dc power supply.

\begin{figure}
    \centering
    \includegraphics[width=0.45\textwidth]{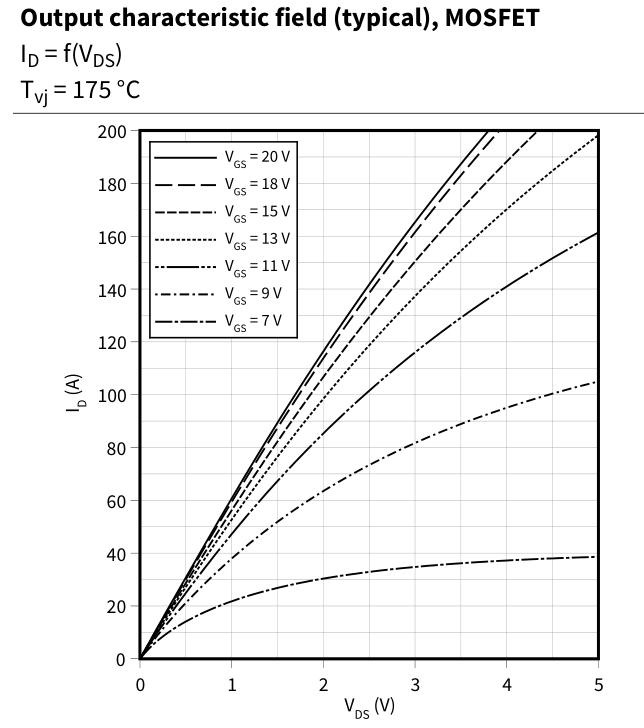}
    \caption{Original output and saturation of FF8MR12W1M1H.}
    \label{fig:datasheet_saturation}
\end{figure}

\begin{figure}
    \centering
    \includegraphics{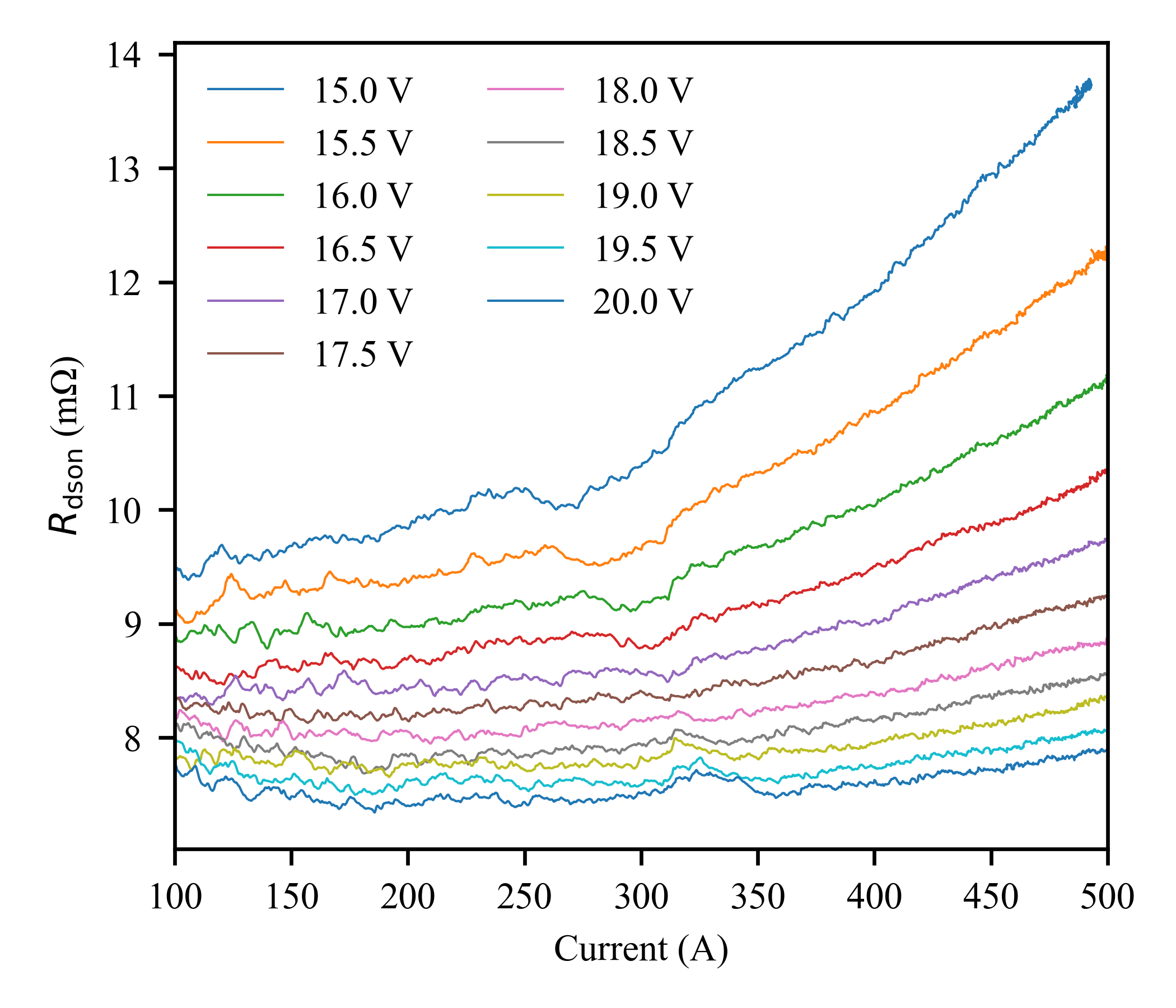}
    \caption{Measured conduction resistance $R_{\text{ds,on}}$ of transistor FF8MR12W1M1H for on-state gate voltages ranging from 15 V to 20 V and a current range up to 500 A.}
    \label{fig:fet_saturation}
\end{figure}

\subsection{Performance and Capabilities of Transistors}
The transistor design is tailored for TMS and systems with a similar transient high-power demand.
Typical TMS pulses last around several hundred microseconds. According to Shannon's sampling theorem, the minimum sampling rate must exceed twice the fundamental frequency of the desired signal and the spectrum that should be practically side-band-free, which requires an effective output bandwidth of at least 50 kHz. Therefore, fast transistors such as MOSFETs are preferred over slow alternatives like IGBTs. 

We chose silicon carbide (SiC) field-effect transistors due to their high voltage capabilities and fast switching dynamics. However, as unpolar devices and due to the typically smaller dies, their  current and particularly their over-load capabilities are lower than for other high-votlage devices, such as insulated-gate bipolar transistors (IGBT). We selected commercially available transistor modules (FF8MR12W1M1H, Infineon Co.) and parallelized units to move the saturation level beyond the TMS requirements (Figure \ref{fig:datasheet_saturation}).

To keep the capacitive parasitics and switching speeds low, we intentionally operate the devices also in the overload range and use a high gate voltage.
Available data on SiC transistors, however, are limited particularly for high currents and gate--source votlages, also for the transistor in question (Fig.\ \ref{fig:datasheet_saturation}). We therefore characterized the modules up to 500~A for gate--source voltages from 15~V to 20~V in steps of 0.5~V (Fig.\ \ref{fig:fet_saturation}). The drain--source resistance stays below the nominal 8~m$\Omega$ in the entire range for gate--source voltages above 19.5~V without any onset of saturation.

Based on these measurements, we set the on-state gate voltage to 20 V to operate the transistors with 500 A peak current. Since this voltage approaches the steady-state limit, additional gate over-voltage protection measures, such as TVS diodes across the gate-source terminals, can be adopted to suppress potential switching voltage overshoot. With the CH2B topology, where two transistors operate in parallel during pulses, we can push the conducting current of each converter module into the kiloampere range.

This experience of overloading transistors beyond their datasheet may also be applied to other applications, particularly in pulsed converters. Conventional applications, such as grid inverters, can be primarily limited by thermal conditions as junction temperature significantly affects saturation behavior. However, in pulsed applications such as TMS, despite the high peak current, the average power remains low. As a result, the junction temperature is less of a concern compared to continuous-load applications.

\section{System Configuration and Operation}
We introduce the system-level operation based on a three-module structure, with the dc power supply connected to the terminal module. 

\subsection{Modulation}
\begin{figure}
    \centering
    \includegraphics{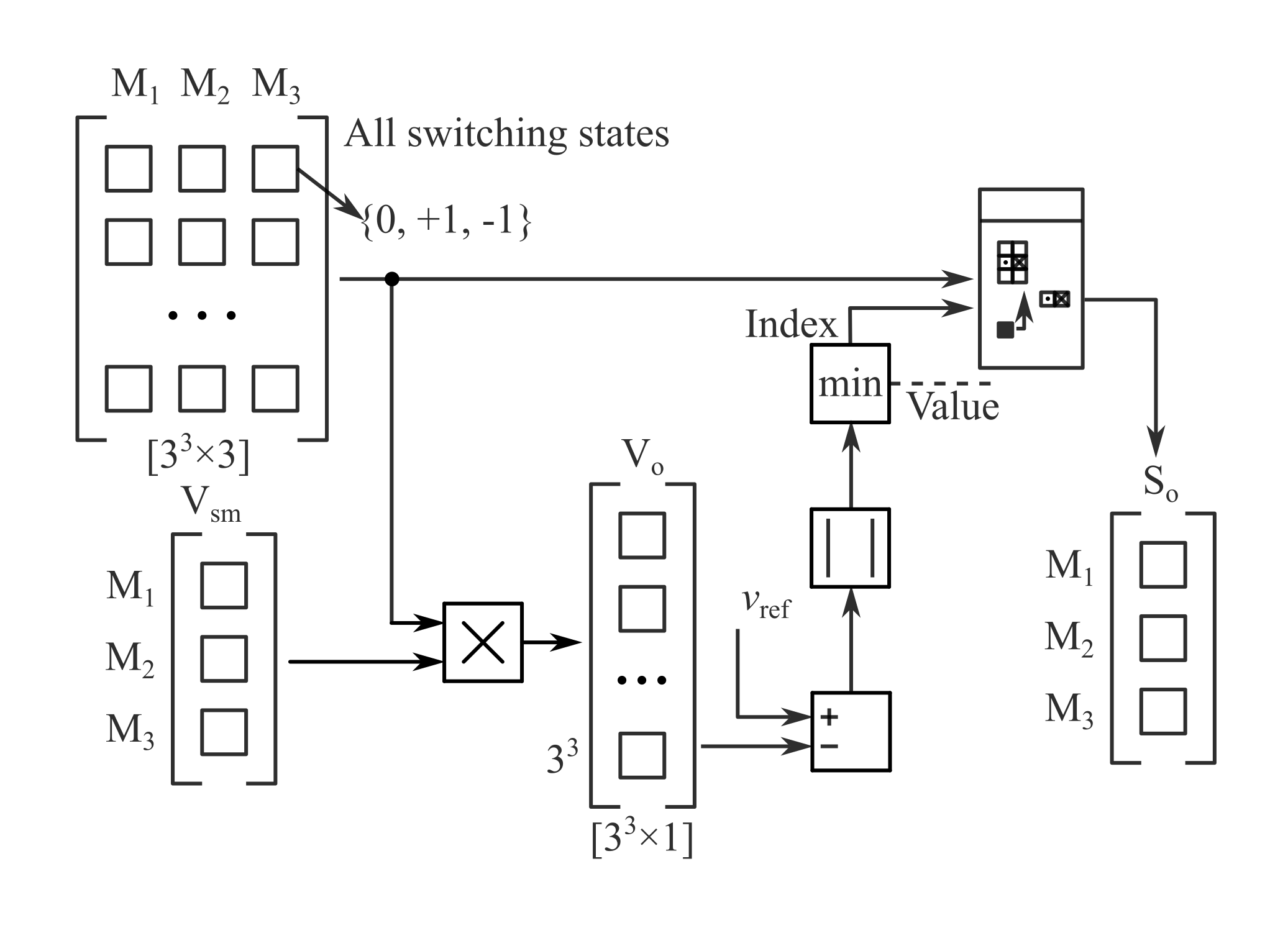}
    \caption{Control diagram of nearest level modulation for a three-module asymmetric multilevel converter.}
    \label{fig:control_diagram}
\end{figure}

With asymmetric module voltage providing a large number of output levels, we prioritize the nearest level modulation (NLM) over carrier-based modulation due to their simplicity and bandwidth advantages.
NLM generates the desired output by approximating the continuous reference signals with discrete output levels. 

Assuming a system consisting of $N$ modules, we can represent its output status with a vector of individual module states per
\begin{equation}
    \overrightarrow{S[k]} = [S_{1}[k],\  S_{2}[k],\ \cdots,\ S_{N}[k]],
\end{equation}
where $S_n[k]$ is the output state of $n^{th}$ module at moment $k$, which follows 
\begin{equation}
    S_n[k] \in \left\{0, +1, -1\right\}.
\end{equation}
The output voltage is obtained as 
\begin{equation}
    v_{o}[k] = \overrightarrow{V}\cdot \overrightarrow{S[k]},
\end{equation}
where $\overrightarrow{V}$ is the array of module voltages and follows 
\begin{equation}
    \overrightarrow{V} = \left[V_1, \ V_2, \ \cdots, \ V_N\right].
\end{equation}

NLM determines the output vector for minimized output deviation as
\begin{equation}
    \overrightarrow{S}[k] = \text{arg }\underset{\overrightarrow{S}}{\text{min }} \left\vert v_{\text{ref}}[k] - v_o[k] \right\vert,
    \label{equ:generic_model_nlm}
\end{equation}
where the denotation $\vert \cdot \vert$ represents the absolute value, $v_{\text{ref}}[k]$ the reference command, while $v_{\text{o}}[k]$ the output voltage.
Figure \ref{fig:control_diagram} illustrates this control scheme and its implementation.

\subsection{Adaptive Optimization of Voltage Asymmetry}
Although asymmetric voltage configurations are previously dominated with binary and ternary setups, recent research  reveals that voltage distribution can be optimized to improve output granularity and practicality \cite{iecon_ammc_general}. Therefore, we optimize the asymmetry of module voltages to achieve the best output quality and a reduced voltage gap, as  
\begin{equation}
    \begin{aligned}
        \overrightarrow{V_{\text{opt}}} = \text{arg }\underset{\overrightarrow{V}}{\text{min }} \left\Vert \overrightarrow{v_o} - \overrightarrow{v_{\text{ref}}} \right\Vert,
    \end{aligned}
    \label{equ:optimization_of_voltage_array}
\end{equation}
where the denotation $\Vert \cdot \Vert$ represents the deviation between two signals -- the output voltage total distortion in this paper. 
The output voltage is obtained based on the modulation algorithm in Figure \ref{fig:control_diagram}, as 
\begin{equation}
    v_o = f_{\text{NLM}}\left(v_{ref}, \overrightarrow{V}\right)
\end{equation}
Additional constraints can be added to the optimization, such as limiting the maximum voltage gap $\Delta V_{\text{max}}$ between modules.
We can further obtain a combined aglorithm to optimize the voltage asymmetry as 
\begin{equation}
    \begin{aligned}
        \overrightarrow{V_{\text{opt}}} \quad = \quad &\text{arg }\underset{\overrightarrow{V}}{\text{min }} \left\Vert \overrightarrow{f_{\text{NLM}}\left(v_{ref}, \overrightarrow{V}\right)} - \overrightarrow{v_{\text{ref}}} \right\Vert,\\
        s.t. \, \quad &\text{max}(\overrightarrow{V}) - \text{min}(\overrightarrow{V}) < \Delta V_{\text{max}}.
    \end{aligned}
    \label{equ:optimization_of_voltage_array}
\end{equation}

\subsection{Switched-Capacitor-Facilitated Charging Mechanism}
\begin{figure}
    \centering
    \includegraphics{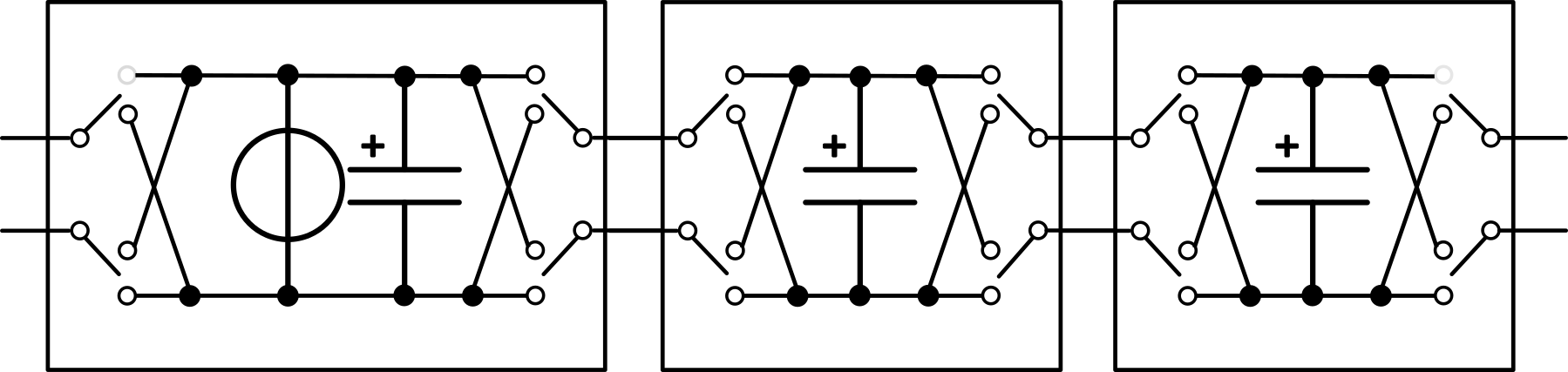}
    \caption{Top-level structure of a three-module prototype.}
    \label{fig:macro_topology}
\end{figure}

\begin{figure}
    \centering
    \includegraphics{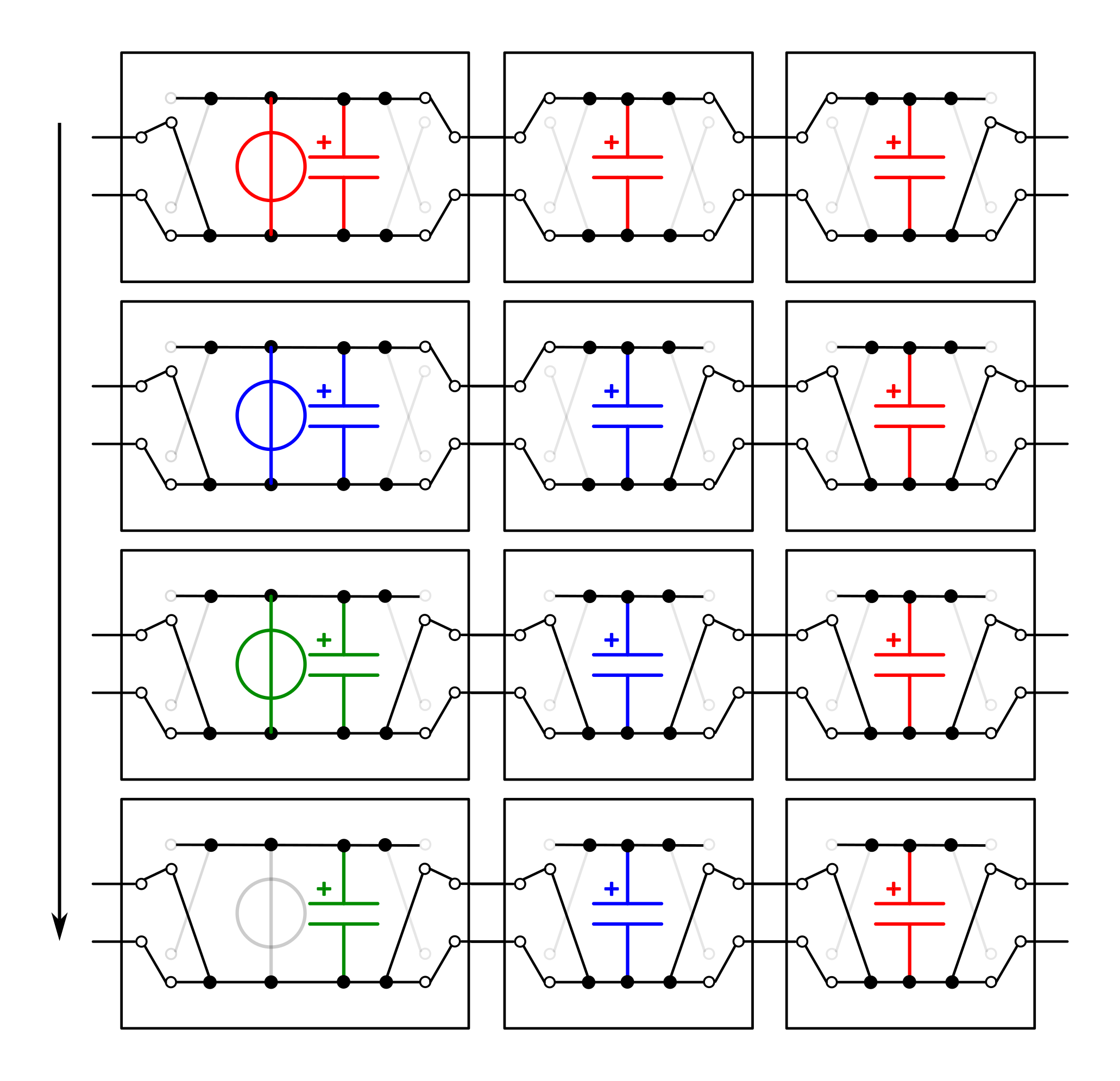}
    \caption{Process of charging modules to different voltages using a single dc power supply.}
    \label{fig:charging_mechanism}
\end{figure}

\begin{figure}
\begin{tikzpicture}[node distance=1cm]
    \node (start) [startstop] {Charging begins};
    \node (parallel) [process, below of=start] {Put all modules into parallel mode};
    \node (v3) [process, below of=parallel] {Regulate the voltage source to $V_3$};
    \node (bypass23) [process, below of=v3] {Switch inter-module connection \#2-\#3 into bypass};
    \node (v2) [process, below of=bypass23] {Regulate the voltage source to $V_2$};
    \node (bypass12) [process, below of=v2] {Switch inter-module connection \#1-\#2 into bypass};
    \node (v1) [process, below of=bypass12] {Regulate the voltage source to $V_1$};
    \node (inhibit) [process, below of=v1] {Inhibit DC power supply (optional)};
    \node (end) [startstop, below of=inhibit] {Charging ends};
    
    \draw [arrow] (start) -- (parallel);
    \draw [arrow] (parallel) -- (v3);
    \draw [arrow] (v3) -- (bypass23);
    \draw [arrow] (bypass23) -- (v2);
    \draw [arrow] (v2) -- (bypass12);
    \draw [arrow] (bypass12) -- (v1);
    \draw [arrow] (v1) -- (inhibit);
    \draw [arrow] (inhibit) -- (end);
\end{tikzpicture}
	\caption{Procedure of charging operation.}
	\label{fig:charging_flow_chart}
\end{figure}
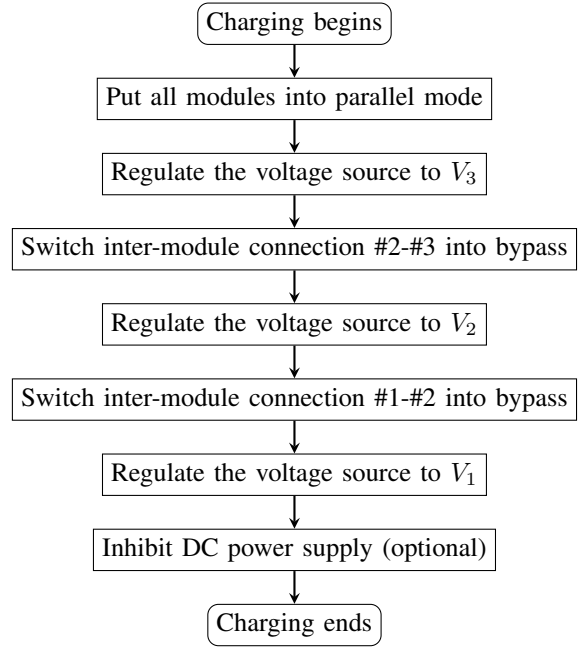

We can simplify the top-level topology of a three-module system to Figure \ref{fig:macro_topology}.
Facilitated by the switched-capacitor feature of the CH2B topology, the proposed hardware solution enables charging multiple modules to different voltages with a single DC power supply. Figures \ref{fig:charging_mechanism} and \ref{fig:charging_flow_chart} summarize the charging process and module operation. With the dc power supply connected to an end module, the charging process begins with all modules in parallel, charging to the desired voltage of the module at the other end. Next, the connection mode between the second-to-last and last modules is switched to bypass, disconnecting the end module from the parallel group. The remaining modules are then charged to the desired voltage of the second-to-last module. This process continues until only the end module that is connected to the dc power supply remains in the parallel group. At this point, all modules have been charged to their respective desired voltages. However, the dc power supply is typically inhibited before firing TMS pulses to prevent measurement artifacts.

\section{Results}
\subsection{Experimental Prototype and Test Platform}
\begin{figure}
    \centering
    \includegraphics{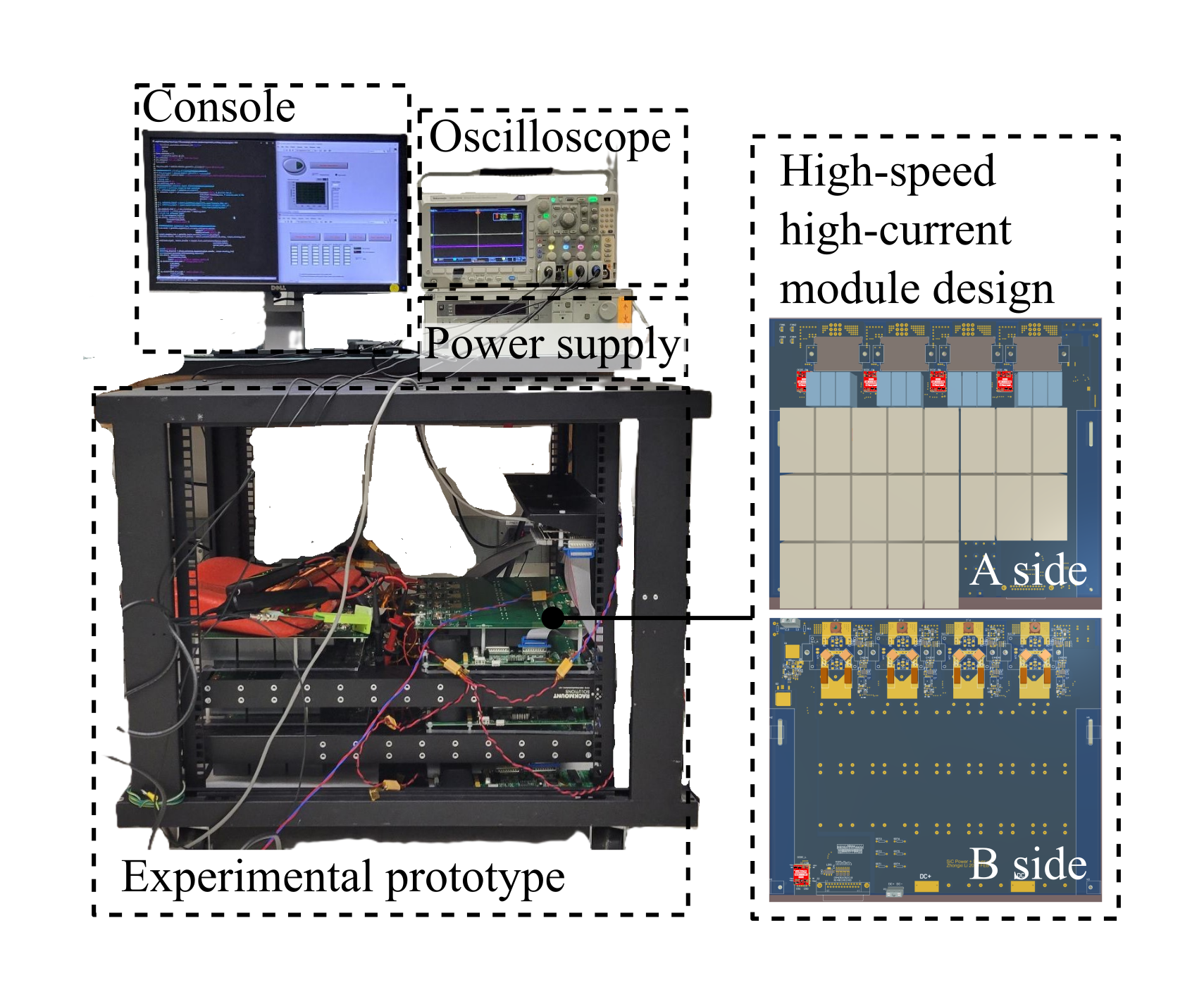}
    \caption{Asymmetric modular pulse synthesizer prototype consisting of three SiC-based cascaded double H-bridge modules and the testing platform.}
    \label{fig:photo_prototype}
\end{figure}

We implemented three experimental setups for a fair head-to-head comparison. Specifically, we built two MPS circuits, one with  three and one with six modules, as well as a three-module AMPS circuit. All prototypes are established with the same modules for consistency, as highlighted in Figure \ref{fig:photo_prototype}. The modules are equipped with SiC MOSFET FF8MR12W1M1H and film capacitors. Each module can handle an output current of at least 1,000~A and switch its transistors at a rate of 50 kHz.

Figure \ref{fig:photo_prototype} shows a three-module asymmetric multilevel prototype and the test platform, including a programmable dc power supply (HP 6030A) and a high sampling rate oscilloscope (MDO3054, 2.5 GSa/s, four channels, Tektronix Co.).

For each symmetric circuit, we explored different modulation methods, including the nearest level modulation (NLM) and phase-shifted carrier (PSC) pulse-width modulation (PWM). For the proposed asymmetric modular pulse synthesizer, we only applied NLM, as it provides 27 output levels, thus a great output resolution. However, we explored two variations of voltage asymmetry. One is a geometric array, which creates a voltage differential ratio of 1.5 between adjacent modules, and the other is a customized voltage array, derived according to the optimization suggested in (\ref{equ:optimization_of_voltage_array}). The optimized voltage configurations for different trials are summarized in Table \ref{tab:voltage_ratios}.
\begin{table}[h]
    \centering
    \caption{Optimized Voltage Asymmetry}
    \renewcommand{\arraystretch}{1.2} 
    \begin{tabular}{l|c|c|c}
        \hline
        Reference Signal & \(V_1\)(\%) & \(V_2\)(\%) & \(V_3\)(\%) \\
        \hline
        Monophasic & 27.8 & 32.0 & 40.2 \\
        Biphasic & 23.4 & 35.0 & 41.6 \\
        Gaussian Polyphasic & 24.3 & 35.1 & 40.6 \\
        \hline
    \end{tabular}
    \label{tab:voltage_ratios}
\end{table}

\subsection{Experimental Results and Comparison to Prior Art}
\begin{figure*}
    \centering
    \includegraphics{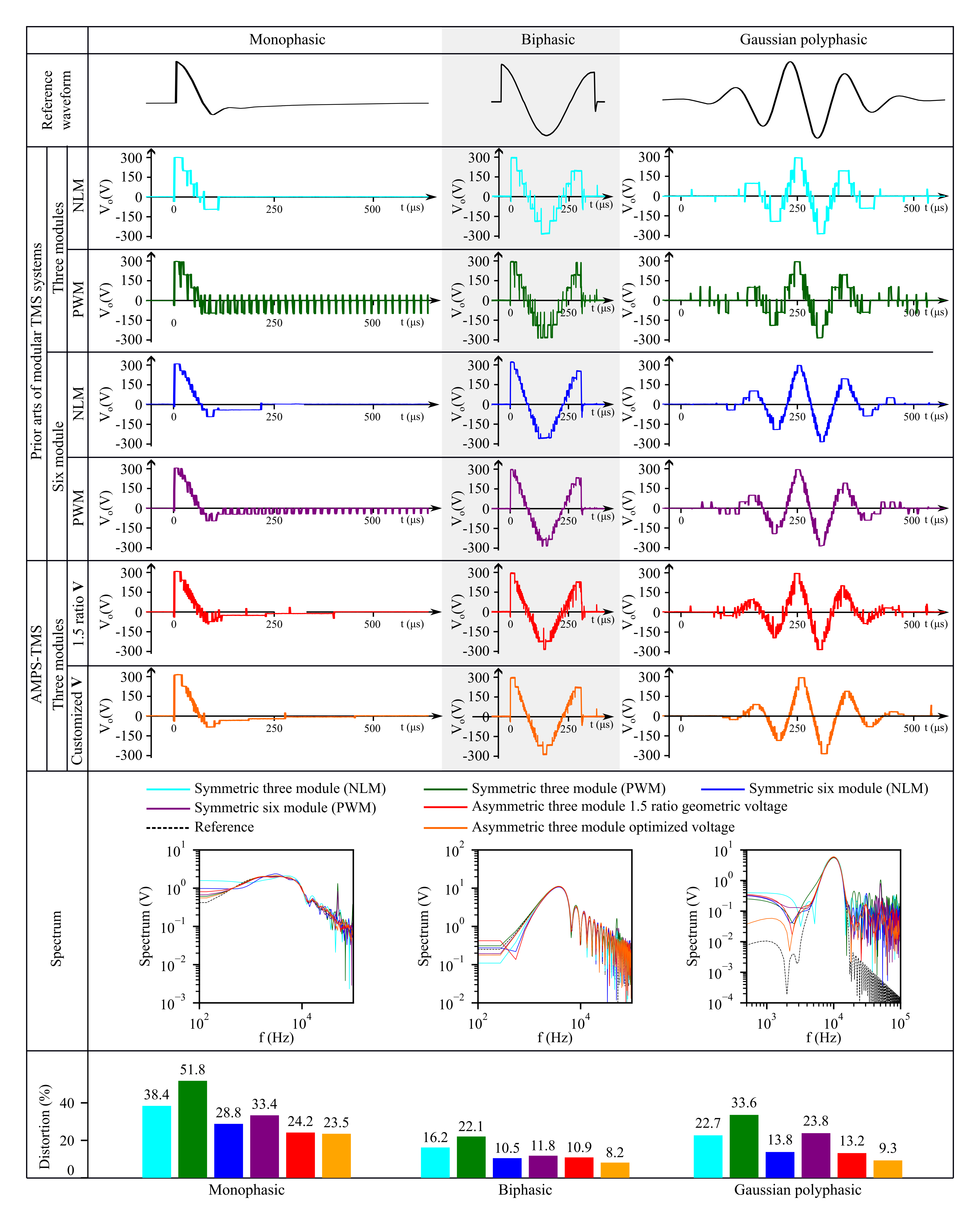}
    \caption{Performance comparison between the suggested asymmetric pulse synthesizer with fixed geometric as well as optimized module voltage sequence and prior-art modular TMS circuits through normalized electric field waveforms, output spectrum, and total distortion.}
    \label{fig:comparison_of_mps_arts}
\end{figure*}
For each TMS implementation, we selected three waveforms to estimate their performance. Two are typical TMS pulses -- monophasic and biphasic -- and the other is a Gaussian polyphasic signal configured with a fundamental frequency of 10 kHz and a standard deviation of $8 \times 10^{-5}$. The Gaussian signal has a smooth starting and ending phases of sinusoidal waveform, thus covers a full range of modulation indices.

We compare the proposed AMPS technology with the prior art with respect to their output voltage waveform, spectrum and total distortion, as shown in Figure \ref{fig:comparison_of_mps_arts}. 
Although with only three modules,  the proposed AMPS circuit exhibits a better performance across all trials, especially with the optimized voltage distribution. 

When equipped with the same number of modules and operated with the NLM approach, the proposed AMPS technology significantly outperforms prior-art symmetric modular circuits. 
Compared to the three-module symmetric circuit, the asymmetric solution reduces the total distortion from 38.4\% to 23.5\% for the monophasic pulse, from 16.2\% to 8.2\% for the biphasic pulse, and from 22.7\% to 9.3\% for the Gaussian polyphasic pulse. Furthermore, the three-module AMPS prototype even has a better performance than the six-module symmetric systems and achieves a 5.3\% lower distortion for the monophasic pulse, 2.4\% for the biphasic pulse, and 4.5\% for the Gaussian polyphasic pulse.

Whether symmetric or asymmetric, all modular circuits are most challenged by the so-called monophasic TMS waveform, which is widely used in brain physiology work. Monophasic pulses cause two to three times higher total distortion than biphasic waveforms. This challenge arises from the long tail of the monophasic pulse, corresponding to the region of low modulation index, where the performance of PWM techniques deteriorates. However, the combination of the NLM approach and a significant number of output levels of asymmetric multilevel converters can produce the output with minimized error without fluctuating between two levels. The fine granularity can manage also shallow transients associated with low-frequency content and explains why the AMPS prototype achieve the greatest distortion reduction in monophasic trials.

Although PWM techniques can generate a visually smoother current waveform, they induce harmonic distortion around 50~kHz, as shown in the spectrums in Figure \ref{fig:comparison_of_mps_arts}. 
As a result, they present a higher total distortion level than the NLM approach. Since neurons exhibit strong nonlinear behavior, linearly separating different spectral components intended to act independently on them is not appropriate. A high output resolution is therefore also required to sufficiently reduce spectral side-bands.

\section{Conclusion}

We proposed a modular pulse synthesizer, which uses an intentional spread in module voltage to increase the number of available output levels for a high resolution. This paper details the module design, including the topology and the potential of the transistors, as well as the system-level structure and operation. Whereas many conventional low-power asymmetric cascaded bridge converters struggle with maintaining the module voltage levels, we introduce a switched-capacitor  charging mechanism. Compared to the prior art, our experimental prototype achieved better output quality, although it uses only half the number of modules.

\bibliographystyle{Bibliography/IEEEtranTIE}
\bibliography{Bibliography/IEEEabrv,Bibliography/refs}\ 

\end{document}